\documentclass[12pt,preprint]{aastex}

\begin{document}
 
\newcommand{\kms}{km s$^{-1}\;$}
\newcommand{\msun}{M_{\odot}}
\newcommand{\rsun}{R_{\odot}}
\newcommand{\rat}{$R_{2}$~}
\newcommand{\fbss}{$F_{BSS}^{HB}$}
\newcommand{\mav}{\langle M_{HB} \rangle}
 
\title{Blue Stragglers in Low-Luminosity Star Clusters\footnote{Based
on observations with the NASA/ESA {\it Hubble Space Telescope},
obtained at the Space Telescope Science Institute, which is operated
by the Association of Universities for Research in Astronomy, Inc.,
under NASA Contract NAS 5-26555.}}

\author{Eric L. Sandquist} 
\affil{San Diego State University, Department of Astronomy, San Diego,
CA 92182}
\email{erics@mintaka.sdsu.edu}

\begin{abstract}
  We examine the blue straggler populations of 13 low-luminosity ($M_{V_t} \ga
  -6$) globular clusters and 2 old open clusters. These clusters test blue
  straggler formation in environments intermediate between higher luminosity
  (and usually higher density) clusters and the Galactic field. The
  anti-correlation between the relative frequency of blue stragglers ($F_{BSS}
  = N_{BSS} / N_{HB}$) and cluster luminosity continues to the lowest
  luminosity clusters, which have frequencies meeting or exceeding that of
  field stars. In addition we find that the anti-correlation between straggler
  frequency and central density disappears for clusters with density less than
  about $300 L_{V,\odot}$ pc$^{-3}$, although this appears to be an artifact
  of the correlation between cluster luminosity and central density.  We argue
  on observational (wide, eccentric binaries containing blue stragglers in
  M67, and the existence of very bright stragglers in most of the clusters in
  our sample) and theoretical grounds that stellar collisions still
  produce a significant fraction of the blue stragglers in low luminosity star
  clusters due to the long-term survival of wide binaries.
\end{abstract}



Blue stragglers are one of several exotic stellar populations whose creation
seems to be intimately tied to the dynamical environment that stars find
themselves in. These stars are probably created when the mass of a normal main
sequence star is increased by external means: the longest-surviving
explanations involve either mass transfer within a binary system or the merger
of stars in a binary or during a collision. The main debate is over which
mechanisms play the largest roles in different environments. In the Galactic
field at one extreme, the probability that two field stars (or binary systems)
will strongly interact after they leave their birth environment is expected to
be vanishingly small. Indeed, radial velocity surveys have identified a high
fraction of single-line spectroscopic binaries with low eccentricity among
field blue stragglers, which supports the idea that these stragglers were
formed via binary mass transfer \citep{ps00,carn01,carn}.

At the other extreme are the most luminous ($M_{V_t} < -6$) globular
clusters. \citet[][hereafter P04]{piotto} found a strong {\it anti-correlation} between
blue straggler frequency [relative to horizontal branch (HB) stars]
and integrated cluster magnitude, and a weaker anti-correlation with
central density. (A similar relation can be seen in \citealt{ferr}.)
Because expected collision rates correlate with cluster mass,
\citet{davies} proposed that primordial binaries that would form
stragglers were being depleted earlier in the history of a massive
cluster, thanks to exchanges of massive stars into the binaries and
subsequent mass transfer.  For the most massive clusters ($M_{V_t} \la
-8.8$), they predicted that star collisions would take over as the
dominant mechanism for forming stragglers.

Regardless of the interpretation of the P04
results, the connection between the field and cluster environments
has not been made. To this end, we have examined low-mass
globular clusters and old open clusters to see whether the straggler
populations still show the influences of the cluster environment, or
whether that influence has begun to disappear.

\section{Observational Material}\label{obs}

To collect the largest possible samples of blue stragglers and
comparison stars for low-luminosity globular clusters, we used 
high quality ground-based photometry from the literature and
photometry from the {\it Hubble Space Telescope (HST)} archives (all
using the WFPC2 camera). In all cases, we chose datasets with little
photometric scatter at the turnoff, so that contamination of the
blue straggler sample by normal main sequence stars was minimal.
All {\it HST} data was reduced in a uniform manner using the
HSTPHOT package \citep{dolphin}.

\begin{figure}
\plotone{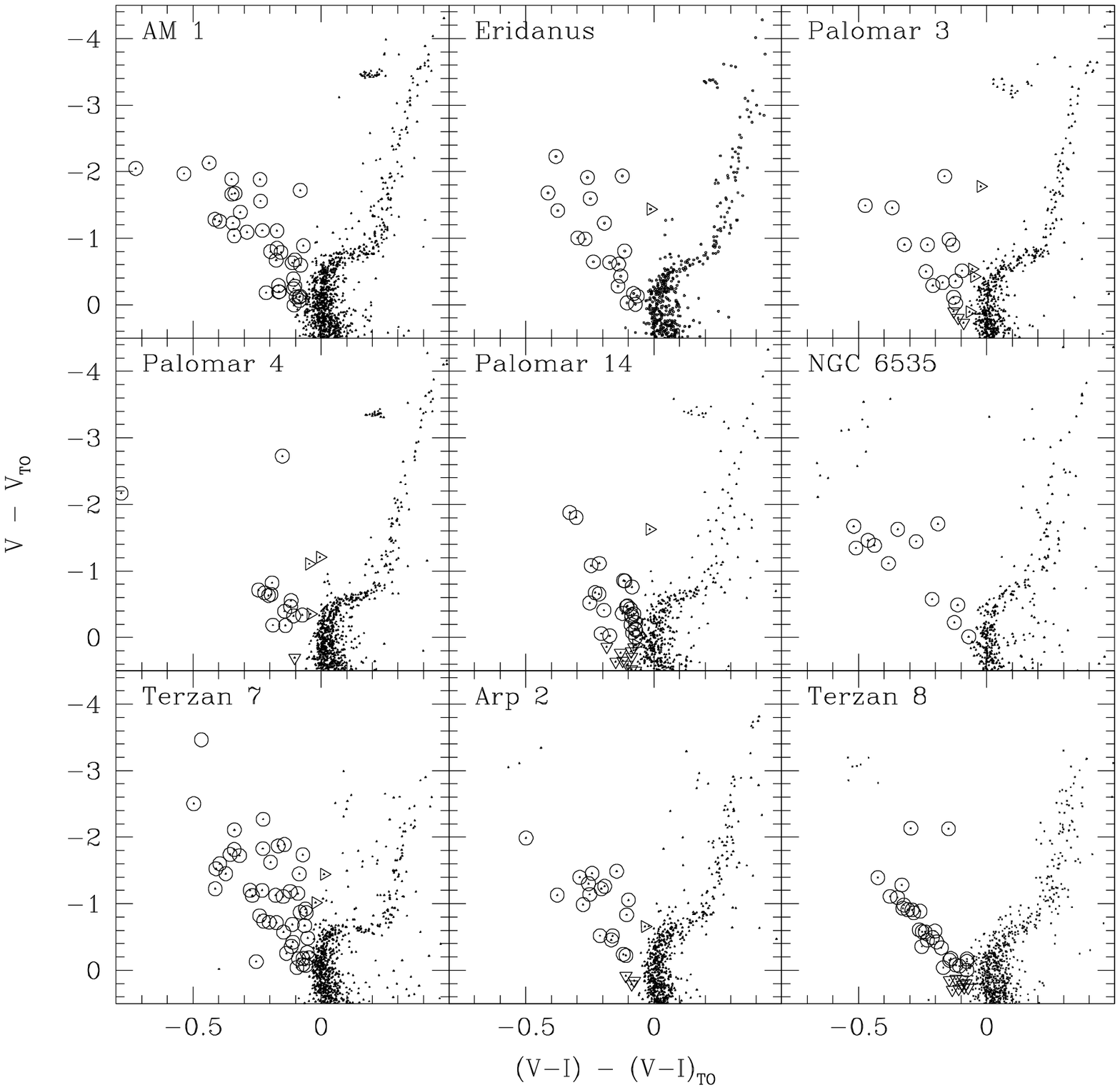}
\caption{CMDs for 9 globular clusters with {\it HST} photometry.
  Open circles are stragglers used in this study, while open triangles show
  candidates too red or too faint for selection.  For several
  of these clusters, additional photometry was used to identify
  other stragglers outside the primary field.\label{cmds}}
\end{figure}

Our final sample consists of 13 globular clusters with $M_{V_t} > -6$ (9 with
HST data). Color-magnitude diagrams (CMDs) for the primary {\it HST} fields
are shown in Fig. \ref{cmds}. The blue stragglers discussed here are all
brighter than the cluster turnoff and bluer (by 0.05 in $(B-V)$), and there
was typically no difficulty with field star contamination of the portion of
the CMD populated by stragglers. Although most of our
straggler identifications are based on $V-I$ colors (as opposed to $B-V$ for
P04), there were few stars for which the choice of color could affect the
selection. In some clusters, a significant field star population redder than
the cluster turnoff and brighter than the subgiant branch made it impractical
to use cluster giants as a comparison population, so we used HB stars instead.
Clusters with large field contamination could sometimes be used if they had
blue HB morphologies. The complete sample of blue stragglers and HB stars will
be presented in a companion paper.


We also examined the blue straggler populations of two well-studied old open
clusters: M67 and NGC 188. Both of these clusters have extensive proper motion
membership information available that allows us to identify stragglers, and to
estimate the absolute magnitude of the cluster by summing the fluxes of member
stars ($M_{V_t} \approx -3.9$ and $-3.8$, respectively).  We have also
estimated the central luminosity densities ($\log \rho_0 \sim 2.3$ and 2.2,
respectively) using structural information from \citet{bb} and \citet{bbs}.
Although these clusters have lower ages (and more luminous stars on
average\footnote{Since cluster luminosity and luminosity density are being
  used as stand-ins for cluster mass or central mass density, the open cluster
  points should be offset toward lower luminosity and density.}), these two
clusters have characteristics comparable to the faintest globular clusters.


P04 used two measures to judge the relative frequency of blue stragglers in a
cluster: the population ratio $F_{BSS}^{HB} = N_{BSS} / N_{HB}$, and the
number of stragglers per absolute visual flux unit $N_S^{BSS}$. In Fig.
\ref{fbss}, we plot \fbss ~ versus absolute cluster magnitude and central
density. Values for these cluster properties were taken from the compilation
of \citet{harris}.  With the inclusion of data for these 13 clusters, we
extend the cluster sample fainter about 3 magnitudes in $M_{V_t}$ and more
than 3 orders of magnitude in central density. The significant anti-correlation
between \fbss ~ and $M_{V_t}$ seen by P04 continues to the low luminosity end
of the sample. That end is anchored by the clusters E 3 ($M_{V_t} = -2.77$)
and Palomar 13 ($M_{V_t} = -3.74$). E 3 has no known HB stars (one unconfirmed
candidate), and so produces a lower limit. The majority of the clusters have
higher relative frequencies than all but a handful of the clusters in the P04
sample. (We also examined the cluster M71, which was the faintest cluster in
the P04 data, to verify that our straggler selection produced a straggler
frequency equal to theirs to within the errors.) None of the clusters examined
here has \fbss $< 0.72$, which is a noticeable departure
from the behavior of more luminous clusters in spite of the
significant variations from cluster to cluster.

Although the open clusters that we have been able to include are
significantly younger than the globular clusters in the sample (4 Gyr
for M67, 6 Gyr for NGC 188), they appear to follow the relation
defined by the globular clusters, and agree with the high value found
for Pal 13. We have used red clump stars (in NGC 188, including
two blue subdwarfs) as the comparison populations for the open
clusters, but this should not be an issue since the lifetimes of stars
in this phase is within about 10\% of the lifetimes of red HB stars in
globular clusters \citep{girardi}. This puts the open cluster
straggler populations in a new context --- normal for their total
masses, not abnormally large.



\section{Relation to Field Blue Stragglers and Open Clusters\label{field}}

Because our sample of low-luminosity clusters also contains some of the lowest
density clusters, we compare to the population of field blue stragglers
\citep{carn,carn01,ps00}. The prevalence of single-lined low-eccentricity
spectroscopic binaries in the field sample implies that stable binary mass
transfer is the main production mechanism among field stars. (Coalescence of
close binaries like W UMa systems would produce the rest.) The
4 lowest-luminosity clusters in our sample are known to have binary fractions
comparable to that of field stars, based on precise photometry of the main
sequence [$f_b > 0.30 \pm 0.04$ for Pal 13 \citep{clark}; $f_b > 0.29 \pm
0.09$ for E3 \citep{veron}; $f_b > 0.38$ for M67 \citep{mmj}; $f_b \ga 0.5$
for NGC 188 \citep{von}]. \citet{ps00} found that low-luminosity ($M_{V_t} >
-6$) globular clusters have larger \fbss ~ values than higher luminosity
clusters and estimated that \fbss $\approx 4$ in the field. 
We find that {\it only}
the four least luminous clusters have BSS frequencies that are consistent with
(Palomar 13, M67, NGC 188) or exceed (E3) the Preston \& Sneden estimate for
the field. 
Thus it
seems that all but the least luminous globular clusters are less efficient at
producing stragglers than the field.

Binary systems that transfer mass are quite ``hard'' systems with binding
energies that are considerably larger than kinetic energies of the stars they
encounter in most clusters \citep{davies}. A figure of merit is $\eta = G m_1
m_2 / a \sigma^2 \langle m \rangle$, where $m_1$, $m_2$, and $\langle m
\rangle$ are masses of the binary components and the average single star, $a$
is the binary separation, and $\sigma$ is the cluster velocity dispersion
\citep{ivan}. The boundary between ``hard'' and ``soft'' binaries is $\eta
\approx 1$, while mass transfer binaries in the clusters described here have
$\eta \ga 100$.  In general, such hard binaries should survive in a cluster
and produce stragglers via mass transfer. Do stellar collisions still produce
stragglers though?

For the nearby open clusters, studies of {\it individual}
blue stragglers provide information on the formation
mechanisms. In NGC 188, 
the straggler PKM 4330 has a high membership probability ($P_\mu = 82$\%;
Platais et al. 2003), but stands more than 3 magnitudes above the turnoff
magnitude, which probably means it has more than twice the turnoff mass and
results from a multi-star collision. There is a much larger body of
information on M67 stars. Like PKM 4330, the brightest straggler
(S977) appears to have a mass more than twice the cluster turnoff mass. There
are many variable blue stragglers, including S1036 (W UMa variable), S1082 (RS
CVn variable, X-ray source, possible triple system containing {\it two} blue
stragglers), and S1280 and S1284 ($\delta$ Scuti pulsators, S1284 in a
short-period eccentric binary).  Several more blue stragglers are in
long-period spectroscopic binaries: S752, S975, and S1195 (low-eccentricity
orbits, S975 with hot companion), and S1267 and S997 (eccentric orbits).
Roughly half of these systems probably formed via normal binary mass transfer
(S752, S975, S1036, and S1195), but an equally large number of systems
probably require a collisional mechanism involving more than 2 stars (most
notably, S977, S997, S1082, S1267, and S1284)\footnote{Because M67 is the
  youngest cluster discussed here ($\sim 4.0$ Gyr; \citealt{vdbstet}), we must
  keep in mind that these stragglers would have evolved and died by the time
  the cluster's age reached that of a globular cluster ($\sim 12$ Gyr).}.

\section{Relation to Higher Luminosity Clusters}

\begin{figure}
\plotone{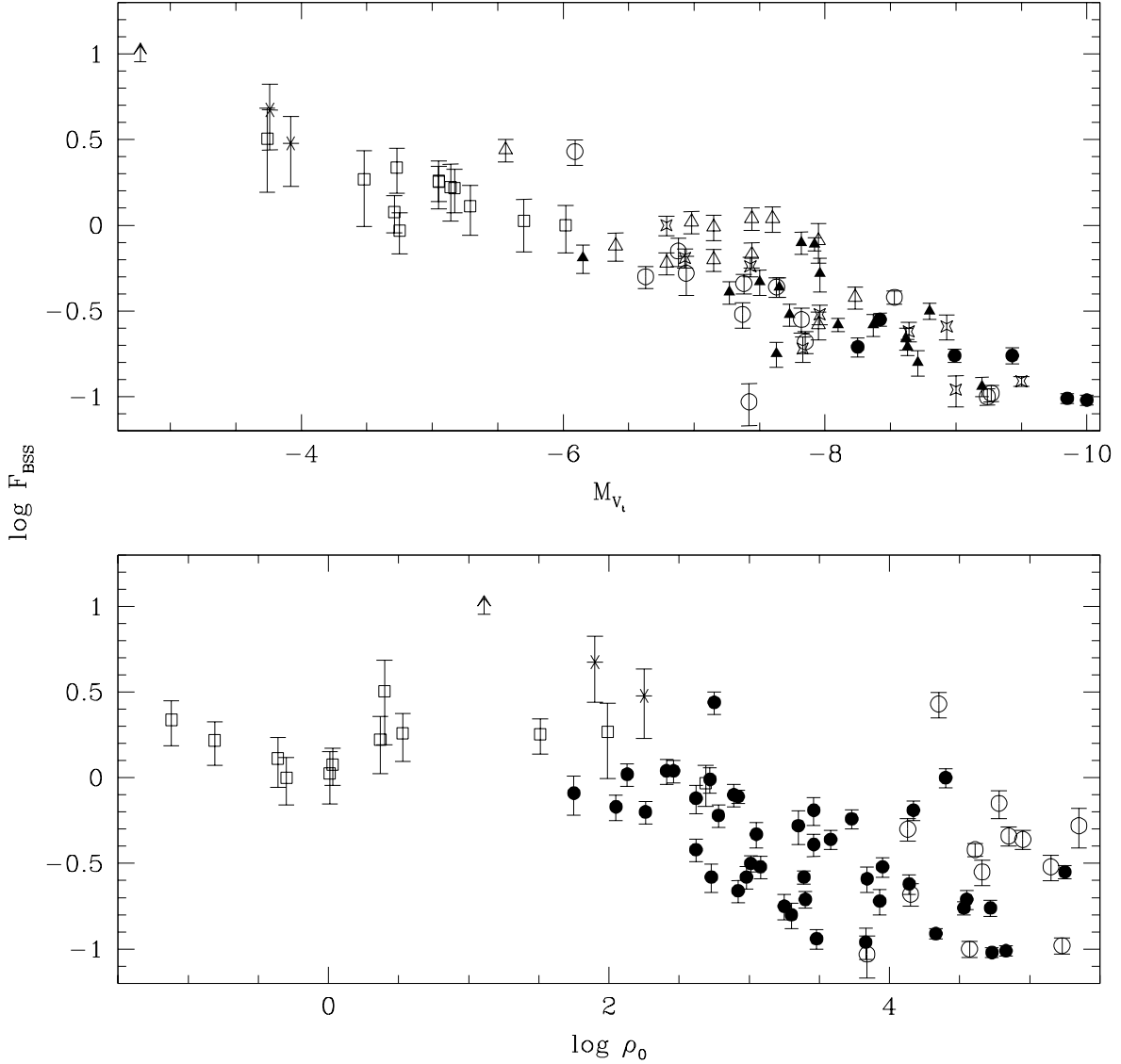}
\caption{Relative frequencies of blue stragglers as a function of
cluster absolute magnitude and central density. Open squares are
globular clusters and asterisks are open clusters from this study,
while all other points are from the {\it HST} Snapshot survey of
\citet{piotto}. Open circles are post-core-collapse clusters. In the
left panel, other symbols represent clusters in different ranges of
central density: $\log \rho_0 < 2.8$: open triangles; $2.8 < \log
\rho_0 < 3.6$: filled triangles; $3.6 < \log \rho_0 < 4.4$: stars;
$\log \rho_0 > 4.4$: filled circles.\label{fbss}}
\end{figure}

\begin{figure}
\plotone{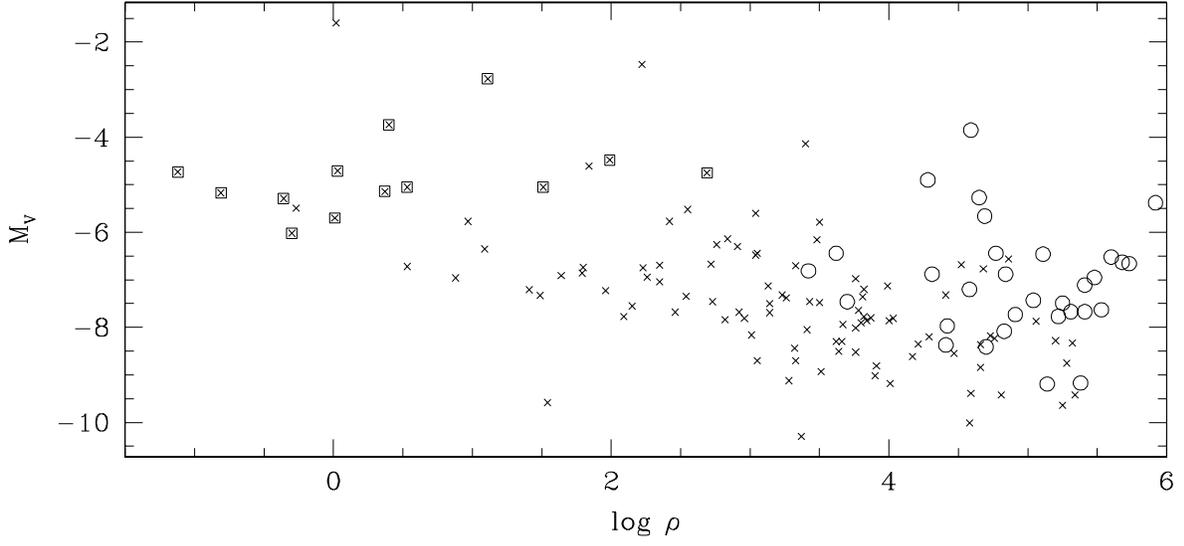}
\caption{Cluster absolute magnitude versus central density for
Galactic globular clusters \citep{harris}. Open circles are
post-core-collapse clusters, and open squares are globular clusters in this
study. Central density is in $L_{V,\odot}$ pc$^{-3}$.\label{pars}}
\end{figure}

The clusters in this study reveal a continuation of the strong
anti-correlation between straggler frequency and cluster luminosity seen by
P04, and a break in the weaker anti-correlation with cluster density (lower
panel of Fig. \ref{fbss}.
An important question is whether the dependence on cluster
density has physical meaning or whether it is an artifact of a correlation
between cluster properties. Fig. \ref{pars} shows the relation between
integrated magnitude and central density \citep{harris}. The similarity of
this plot and the bottom panel of Fig. \ref{fbss} leads us to believe that the
density relationship is an artifact --- the clusters E3 and Pal 13 most
clearly stand out as clusters with high straggler frequencies, low luminosity,
and relatively high central densities. Low-luminosity clusters ($-6 < M_{V_t}
< -4$) cover almost the entire range of central densities, while straggler
frequencies remain uniformly high. We must keep in mind that for most of the
clusters surveyed by P04 the blue straggler samples are mostly
confined to the core, and selection effects in the P04
cluster sample (aside from $M_{V_t} > -6$) may influence the
interpretation of these figures. However, central density is not as good a
predictor of the size of the straggler population as integrated cluster
luminosity.


Does this mean that strong gravitational interactions between stars are
unimportant in the production of stragglers in low-luminosity clusters? As
mentioned in \S \ref{field}, a significant fraction of stragglers in M67 are
in binaries showing evidence of strong multiple-star interactions.  Stars at
the bright end of the straggler luminosity function should also reflect the
relative importance of stellar collisions and binary coalescences since mass
transfer is very unlikely to be 100\% efficient during the formation of a
straggler in a wide binary. As shown in Fig. \ref{cmds}, the clusters in our
low-luminosity sample still produced a substantial fraction of
stragglers at the high luminosity end (more than 10\% have $V - V_{TO} <
-1.75$). 9 of the 13 globular clusters (as well as M71) and both open clusters
have at least one straggler more than 2 magnitudes brighter than the turnoff,
implying a collision of at least three stars.


We can attempt to qualitatively understand the problem by
examining the timescale for collisions $t_{coll}$
\citep{ivan,davies}. The dominant factor in $t_{coll}$ is number density
($t_{coll} \propto n^{-1}$). Our cluster sample covers more than three orders
of magnitude in central density, and when combined with the sample of P04, the
total range is more than six orders of magnitude. So, it might be tempting to
believe that the near constancy of \fbss ~ for many clusters with $\log \rho_0
\la 2.5$ is an indication that primordial binary star mechanisms dominate and
strong interactions have become unimportant.


The velocity dispersion $\sigma$ is also a few times lower for
low-luminosity clusters compared to most clusters in the P04 sample:
from a maximum of 2.4 \kms for NGC 6535 \citep{pm} to $0.6
- 0.9$ \kms for Palomar 13 \citep{blech} and 0.8 \kms for M67 \citep{girard},
to $0.1 - 0.4$ \kms for Palomar 5 \citep{oden}. Low encounter velocities
increase the time between interactions but more strongly increase the
probability of collision via gravitational focusing, so that $t_{coll} \propto
\sigma$ \citep{binn}. Velocity dispersion only varies by a factor of about
100 among all Milky Way globular clusters, so its effects on the 
timescale are usually overwhelmed by those of cluster density.


The velocity dispersion has a larger effect on the statistical properties of
the binary star population though. As velocity dispersion decreases, wider
binaries qualify as ``hard'' ($\eta \propto \sigma^{-2}$). For $\sigma = 10$
km s$^{-1}$, the hard-soft boundary is already at several AU. For most
globular clusters (and all of the ones in the present sample), binary systems
that would ultimately produce blue stragglers via mass transfer in a giant
evolutionary phase are likely to survive due to their hardness, even though
collisions are fairly frequent. For the moment, we only consider
collisions (i.e. strong gravitational interactions) involving binaries, and
not physical collisions of stars.  In a cluster like NGC 6535
at the bright end of our sample,
the collision timescale for the largest binaries that transfer mass in the
giant phase implies that they have a few collisions in the age of the
cluster.  
(Keep in mind that these collision timescales use central
densities, and so are overestimates for the cluster as a whole.)

For lower luminosity clusters, there are competing effects. Lower
velocity dispersion allows wider binaries to survive to the
present day. Photometric measurements of large binary fractions among
some of the low-luminosity clusters in our sample
\citep{clark,veron,mmj,von} support this. Lower cluster density
increases the collision timescale for binaries of a given size,
although this is partly offset by lower $\sigma$ and increased
gravitational focussing. However, the changes in the nature of the
binary population (increased binary fraction and increased average
orbital separation) help enhance the overall binary collision
rate.

The collision timescale for binaries at the hard-soft boundary has
$t_{coll} \propto n^{-1} \sigma^3$, which allows lower velocity
dispersion to partially cancel the effects of lower number density. If
we assume that the present-day binary period distribution is uniform
in $\log_{10} P$ between a minimum period (set by star sizes) and the
hard-soft boundary, then the typical binary involved in a strong
interaction will have $a \approx a_{hs} / \ln \left(a_{hs} /
a_{min}\right)$, where $a_{min}$ and $a_{hs}$ are orbital separations
of the hardest binaries and the hard-soft boundary, respectively. Thus, the
collision timescale for these systems should only be a factor of $5 -
15$ more than the collision timescale for binary systems at the
hard-soft boundary.


A strong interaction involving a close binary is very likely to produce at
least one physical collision of stars ($\sigma_{coll} \sim \pi (a_1 + a_2)^2
(v_c / v_{\infty})^2$, where $v_c$ is the critical velocity needed to unbind
the binary; Fregeau et al. 2004). However, for wide but hard binaries in
low-mass clusters, the main sequence star components present small targets
compared to the binary. For a ratio of stellar radius to orbital separation $R
/ a \la 0.001$, the cross section for collisions decreases approximately as
$(R / a)^{0.75}$, according to Figure 9 of \citet{freg}.
Because the binary collision timescale $t_{coll} \propto a^{-1}$, the
timescale for stellar collisions involving binaries then {\it
decreases} slowly with increasing $a$.

\section{Conclusions}

A combination of effects 
may therefore conspire to produce a significant number of stellar collisions
involving binaries even within low luminosity clusters. The lack of a stronger
correlation between \fbss ~ and central density may result from the effects of
changing velocity dispersion and binary population. This might also partially
explain the apparent lack of correlation between \fbss ~ and the probability
$\Gamma_\ast$ that a given star will have a collision in one year
(P04) since the calculation of $\Gamma_\ast$ from King models does
not account for changes in the binary star distributions between clusters.


Populations of low-luminosity X-ray sources in globular clusters [such as
quiescent low-mass X-ray binaries (qLMXBs), cataclysmic variables, and
millisecond pulsars] appear to behave differently. \citet{pool} found strong
correlations between number of sources and normalized encounter rate $\Gamma$,
and \citet{heink} identified probable qLMXBs that are consistent with
predictions based on tidal capture or exchanges during multiple star
interactions. It is not clear that this is consistent with the scenario of
\citet{davies} in which massive stars exchange into previously-existing
binaries in the most massive clusters, forming stragglers that die off before
the present day. The blue straggler and low-luminosity X-ray populations may
be giving us complimentary information about the importance of collisions and
the prevalence of binaries in star clusters.





We suggest two directions to disentangle the important
factors driving the dynamical evolution of clusters. First, the binary star
populations of more clusters need to be observationally characterized ---
particularly the binary fraction and the period distribution. Second,
models of cluster evolution that examine the production of
stragglers and X-ray sources are needed. The recent hybrid
models (binary population synthesis with dynamics) of \citet{ivan}
provide some guidance, although they modelled clusters that are denser and
more massive than the ones in our sample. They find that the present-day core
binary fraction is inversely related to cluster density, and that the binary
period distribution remains nearly flat in $\log P$ up to periods of $10^4$
d and more for their lowest-density clusters.  The low-luminosity
clusters examined in this study make perfect subjects for numerical studies of
cluster dynamics with current generations of computer codes.

\acknowledgments We would like to thank the anonymous referee for helpful
suggestions, K. Cudworth, D. Dinescu, B. Mochejska, A. Rosenberg, and A.
Sarajedini for copies of their datasets, and useful
conversations with O. De Marco and J. Fregeau. This work has been funded
through grant AST 00-98696 from the National Science Foundation to E.L.S. and
M. Bolte.

\end{document}